\documentclass{kapproc} % Computer Modern font calls

\usepackage{graphicx}

\normallatexbib

\begin{document}

\articletitle{Quantum breathers in an attractive fermionic Hubbard
model}

\articlesubtitle{Quantum breathers in a Hubbard model}

\author{J.C Eilbeck}
\affil{Department of Mathematics, Heriot-Watt University\\
  Riccarton, Edinburgh, EH14 4AS, UK} \email{J.C.Eilbeck@hw.ac.uk}

\author{F. Palmero}
\affil{Nonlinear Physics Group. Departamento de
  F\'{\i}sica Aplicada I. ETSI Inform\'atica, Universidad de Sevilla\\
  Avda Reina Mercedes s/n, 41012 Sevilla, Spain} \email{palmero@us.es}

\begin{abstract}
  In one--dimensional translationally invariant anharmonic lattices,
  an extended Bloch state with two or more strongly correlated
  particles is usually called a quantum breather. Here we study an
  attractive fermionic Hubbard model with two kind of particles of
  opposite spin. We discuss the existence of breathers, and several
  effects that break the translational symmetry of the system and
  localize the breather in the lattice.
\end{abstract}

\begin{keywords}
Anharmonic quantum lattices, Quantum breathers, Quantum lattice
solitons
\end{keywords}

\section{Introduction}
Recent theoretical developments and improved experimental techniques
has led to growing interest in the phenomenon of localization of
energy by nonlinearity in anharmonic lattices. The existence and
properties of these intrinsic localized modes, known as discrete
breathers, have been subject of an much investigation (see, e.g
\cite{Brerev} for a number of recent reviews of this topic). At
present, discrete breathers in classical systems is a relatively well
understood phenomenon, but knowledge of the quantum equivalent of
discrete breathers is not very developed.  In particular we restrict
ourselves to a study of small lattices and a small number of quanta
where some numerically exact solutions can be found.  Although of less
interest to the study of bulk matter, such studies are relevant to the
recent developments in quantum nanotechnology and applications in
quantum computing \cite{li03}.

The quantum equivalent of a discrete breather in a translationally
invariant anharmonic lattice is an extended Bloch state with two
or more particles in a strongly correlated state. There exist some
theoretical results (i.e.\ \cite{seg94,Qtrev}), and some
experimental observations of these states in different quantum
systems, as mixtures of 4--methyl--pyridine \cite{Fil}, in Cu
benzoates \cite{Asa00}, and in doped alkalihalides \cite{Sch02}.

Here we present some results on a quantum one--dimensional lattice
problem with a small number of quanta. We study a periodic lattice
with $f$ sites containing fermions, described by an attractive
fermionic Hubbard model (FH) with two kinds of particles with
opposite spins. It is a model of interest in connection with the
theory of high-$T_c$ superconductivity \cite{Mic92}, and it can be
used to describe bound states of electron and holes in some
nanostructures as nanorings (excitons) \cite{Rom95}. Many of the
results could be extended to a great variety of systems,  i.e., we
have obtained similar results with a periodic lattice containing
bosons and described by the quantum discrete nonlinear
Schr\"{o}dinger equation \cite{Eil03}.

This paper is organized as follows: In the next section we present the
model, and in Section 3 we study the existence of breathers in the
simplest nontrivial case. In Section 4, we consider some modifications
that break the translational symmetry of the lattice, and can localize
the breather in the lattice. In Section 5 we extend the previous
results, obtained in the simplest nontrivial case, to more complicated
situations. Finally, in Section 6, we summarize our findings and
present our conclusions.

\section{The model}
We consider an anharmonic lattice with $f$ sites and two kinds of
fermions with opposite spins described by an attractive fermionic
Hubbard model (FH). The Hamiltonian of the system is given by

\begin{equation}
\hat H=-\sum_{j=1}^{f} \gamma_j a_j^\dag a_j b_j^\dag
b_j+\epsilon_ja_j^\dag(a_{j-1}+a_{j+1})+m_e \epsilon_j
b_j^\dag(b_{j-1}+b_{j+1}), \label{Ham_fer}
\end{equation}
where $a_j^\dag (a_j)$ and $b_j^\dag (b_j)$ are raising (lowering)
operators for different electronic spin states, satisfying the
standard fermionic anticommutation relations. The parameter ratio
$\gamma_j/\epsilon_j$ represent the ratios of anharmonicity to
nearest--neighbor, hopping energy, and $m_e$ is the ratio of the
effective mass of one type of fermion to the other. To eliminate the
effects related to the finite size of the chain, we consider
periodic boundary conditions and, initially, a translational
invariant lattice, $\gamma_j=\gamma$ and $\epsilon_j=\epsilon$,
independent of $j$. In general we consider $\epsilon=1$.

The Hamiltonian (\ref{Ham_fer}) conserves the number of quanta
$N$, and it is possible to apply the number--state--method to
calculate the eigenvalues and eigenvectors of the Hamiltonian
operator \cite{sc99}. We use a number--state--basis $
|\psi_n\rangle =[n^a_1,n^a_2,...,n^a_f;n^b_1,n^b_2,...,n^b_f]$, where
$n^a_i$ ($n^b_i$) represents the number of quanta of fermions $a$
($b$) at site $i$. In this case, $N_a=\sum_i n^a_i$, $N_b=\sum_i
n^b_i$, and $N=N_a+N_b$. A general wave function is
$|\Psi_n\rangle =\sum_n c_n|\psi_n\rangle $. As a first step, we restrict
ourselves to study the simplest nontrivial case $N_a=1$, $N_b=1$,
and as a second step we consider more complicated situations with
a small number of quanta, although many of the results are valid
for larger values of $N_a$ and $N_b$. The bound states correspond
to exciton states, localized electron/hole states that may
appear in nanorings.

\section{Quantum breathers in a translational invariant lattice}
In a homogeneous quantum lattice with periodic boundary
conditions, it is possible to block--diagonalize the Hamiltonian
operator using eigenfunctions of the translation operator $\hat T$
defined as $\hat T b_j^\dag = b_{j+1}^\dag \hat T$ ($\hat T
a_j^\dag = a_{j+1}^\dag \hat T$). In each block, the
eigenfunctions have a fixed value of the momentum $k$, with
$\tau=\exp(i k)$ being an eigenvalue of the translation operator
\cite{sc99}. In this way, it is possible to calculate the
dispersion relation $E(k)$ with a minimal computational effort.
The corresponding matrix in the case $N_a=N_b=1$ is
$$
H_k=-\left[
\begin{array}{cccccc}
\gamma & q^* & 0 & . & . & q \\
q & 0 & q^* & 0& . & 0 \\
0 & q & 0 & q^* & . & . \\
. & . & . & . & . & . \\
. & . & . & q & 0 & q^* \\
q^* & . & . & . & q & 0 \end{array} \right],
$$
where $q=(m_e+\tau^*)$.

\begin{figure}[ht]
\begin{center}
 \includegraphics[scale=0.6]{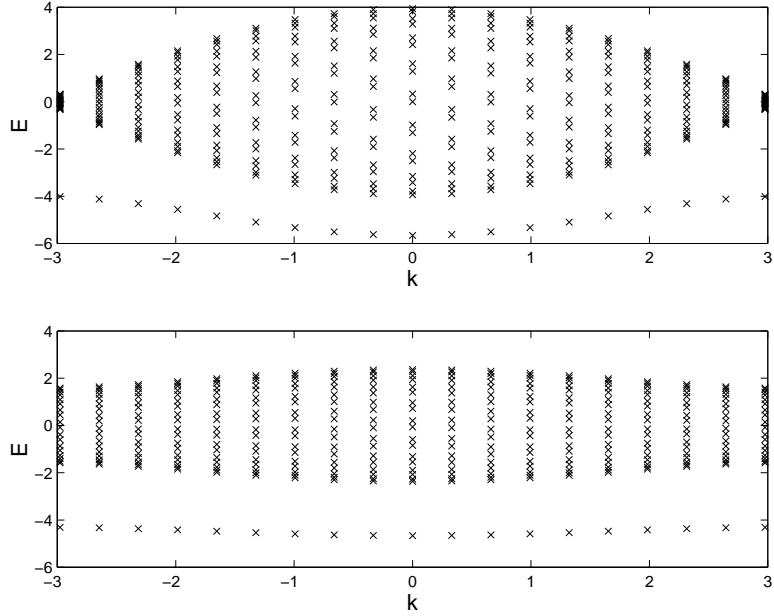}
 \end{center}
\caption{Eigenvalues $E(k)$. $N=2$, $f=19$, $\gamma=4$. $m_e=1$
(top) and $m_e=0.2$ (bottom).}
  \label{fig1}
\end{figure}

In this simplest non--trivial case, if the anharmonicity parameter
is large enough, as Fig \ref{fig1} shows, there exists an isolated
eigenvalue for each $k$ which corresponds to a localized
eigenfunction, in the sense that there is a high probability for
finding the two quanta at the same site. But due to the
translational invariance of the system, there is an equal
probability for finding these two quanta at any site of the
system. In these cases, some analytical expressions can be
obtained in some asymptotic limits (for a recent discussion see
\cite{seg94, sc99, Ei02}). Note that, qualitatively, the existence
of this localized state is independent of the value of parameter
$m_e$.

If we consider for simplicity the case $k=0$, the ground state
unnormalized eigenfunction is
$$
|\Psi\rangle=[10\dots0;10\dots0]+[01\dots0;01\dots0]+
\dots[0\dots01;0\dots01]+O(\gamma^{-1}),
$$
i.e. on a lattice of length $f$, the unnormalized coefficients
$c_i$ of the first $f$ terms are equal to unity and the rest are
$O( \gamma^{-1})$.

\section{Trapping in a lattice with broken translational symmetry }

In this section we will consider some modifications that can break
the translational invariance of the lattice, changing the
coefficients $c_i$ and localize the breather around a particular
point of the lattice. In these cases, the Hamiltonian operator
cannot be block-diagonalized using eigenvectors of the translation
operator. Although the computational effort increases, it is still
possible to calculate its eigenvalues and eigenvectors if $f$ and
$N$ are small enough, by using algebraic manipulation methods and
numerical eigenvalue solvers. In this section we restrict to the
situation $N_a=N_b=1$.

Perhaps the simplest way to break the translational invariance of the
lattice is by considering non--flux boundary conditions to simulate a
finite--size chain. In this case, the solution becomes weakly
localized around the middle of the lattice. If $f$ is high
enough, and we do no take into account boundary effects, this case
reduces to the homogeneous lattice case.

A alternative mechanism for breaking the translational invariance can
be the existence of local inhomogeneities or impurities. In our model,
this can be modeled by making one or more of the $\gamma_j$ or the
$\epsilon_j$ dependent on $j$. This can occur because of localized
impurities or long--range interaction between non nearest-neighbors
sites due to non--uniform geometries of the lattice chain. The
interplay between these two sources of localization, nonlinearity and
impurities is important to understand the properties of these bound
states.

\begin{figure}[h]
  \begin{center}
    \includegraphics[scale=0.25]{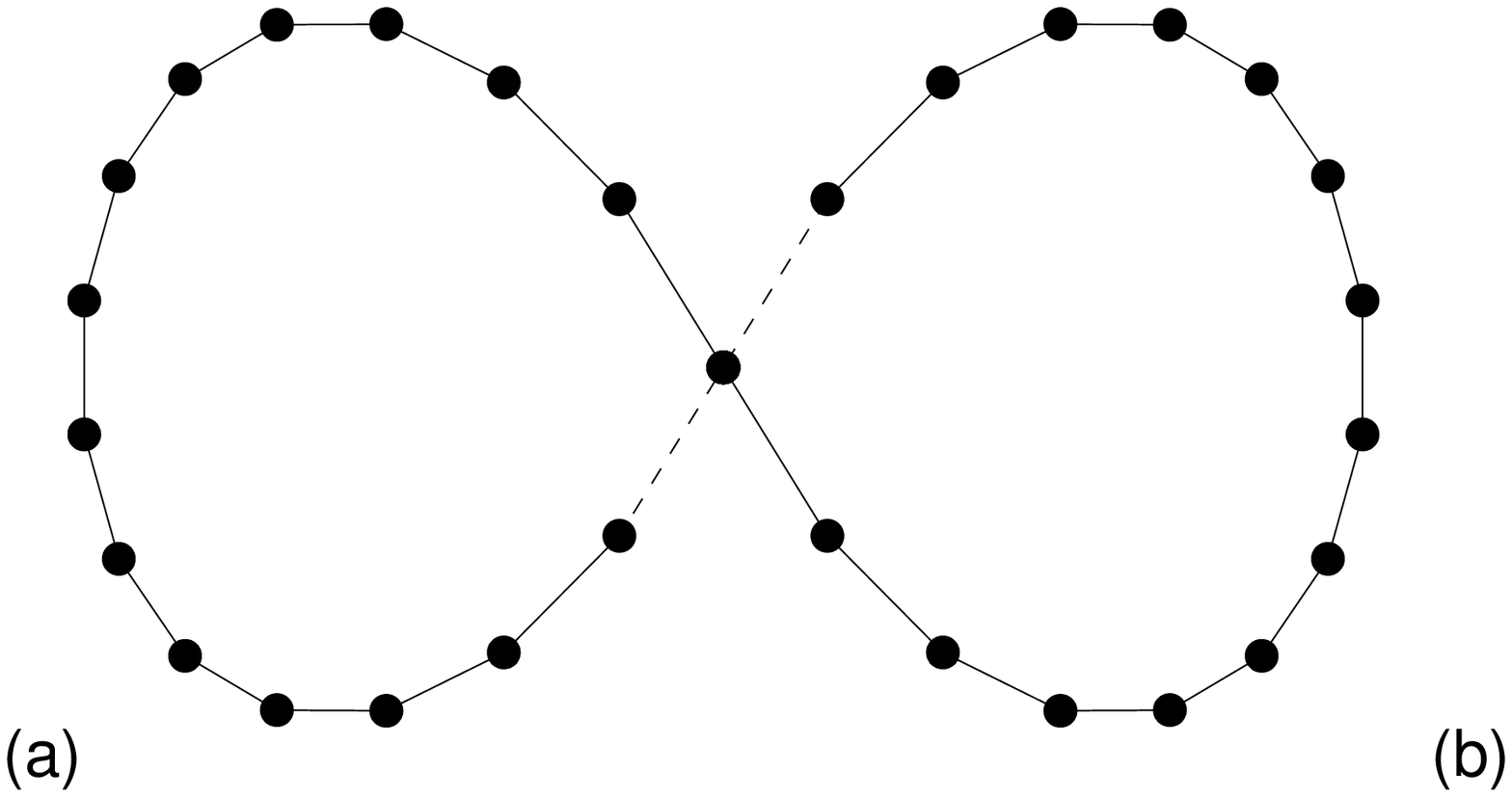}
    \includegraphics[scale=0.3]{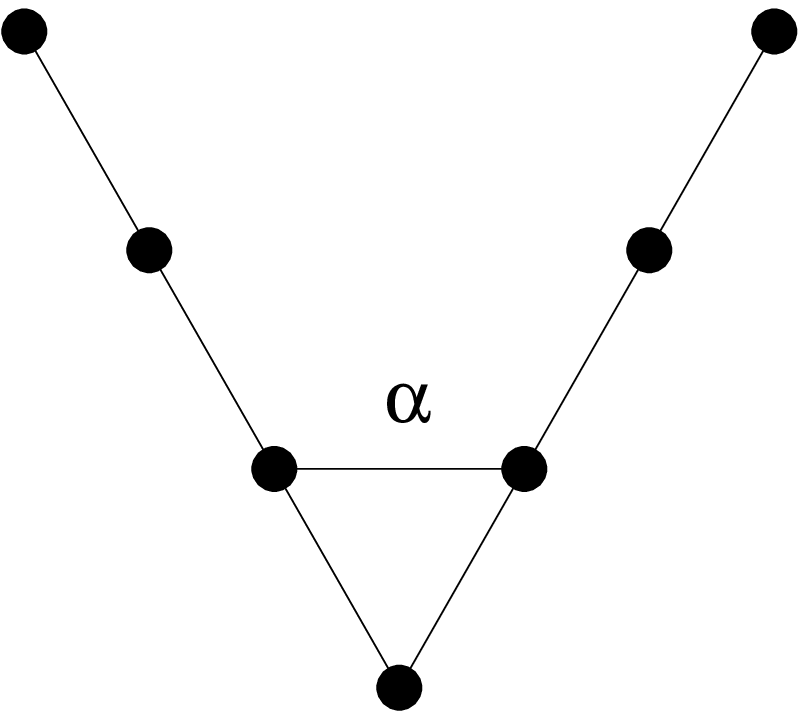}
  \end{center}
  \caption{Two non-uniform chain geometries.}
  \label{fig2}
\end{figure}

Two examples of non--uniform geometries are shown in Fig.\ \ref{fig2}.  In
Fig.\ \ref{fig2}a, a twisted circular geometry causes an interaction
between two sites of the chain, which are distant with respect to
measurement along the length of the chain. This model has been used in
a classical model of a globular protein \cite{ei86}, and it has been
shown that moving breathers described by the DNLS equation can be
trapped at the cross--over point. Fig.\ \ref{fig2}b shows another possible
geometry, a bent chain, that has been recently studied in the context
of the DNLS equation and photonic crystal context \cite{Kiv03} and in
Klein--Gordon systems \cite{Cue03}. In all these cases, the geometry
effects can be modeled by adding a long--range interaction term of the
form
\begin{equation}
\alpha_{\ell,m}(b_\ell^\dagger b_m+b_m^\dagger b_\ell),
\label{lrterm}
\end{equation}
where $\ell$ and $m$ are the neighbouring sites put brought closer in the
twisted--chain case, and $m=m_0-1$ and $\ell=m_0+1$ in the
bent--chain case, where $m_0$ is the vertex of the chain.

We will analyze in more detail these modifications that break the
translational invariance of the system.

\subsection{Localization in a chain with impurities}
We introduce a local inhomogeneity in the anharmonic parameter in our
system and retain periodic boundary conditions, in order to isolate
the effect caused by this local inhomogeneity alone. We put
$\gamma_\ell=\gamma_{imp}$, and $\gamma_j=\gamma$ for $j \neq l$.

\begin{figure}[h]
  \begin{center}
    \includegraphics[scale=0.5]{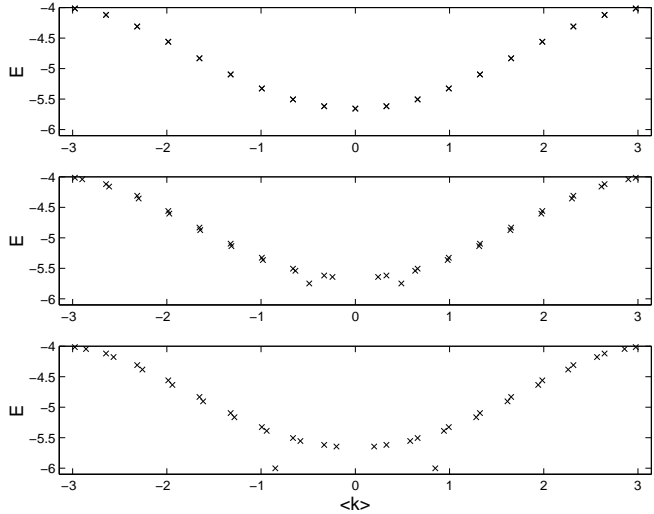}
  \end{center}
  \caption{Eigenvalues E as a function of the expected value of $k$
    corresponding to the localized eigenfunctions. $N=2$, $f=19$,
    $m_e=1$ and $\gamma=4$. Point impurity at the site $\ell=10$.
    Homogeneous chain (top).  $\gamma_{imp}=4.5$ (center).
    $\gamma_{imp}=5$ (bottom).}
  \label{fig3}
\end{figure}

In the homogeneous system, as discussed above, if the anharmonicity
parameter is large enough there exists a high probability of finding
the two particles at the {\em same} site of the chain, but with equal
probability at {\em any} site of the chain.  If we consider a point
impurity, a isolated localized bound state appears, as shown in Fig.
\ref{fig3}. This state has minimal energy and corresponds to the
ground state.

\begin{figure}[h]
  \begin{center}
    \includegraphics[scale=0.3]{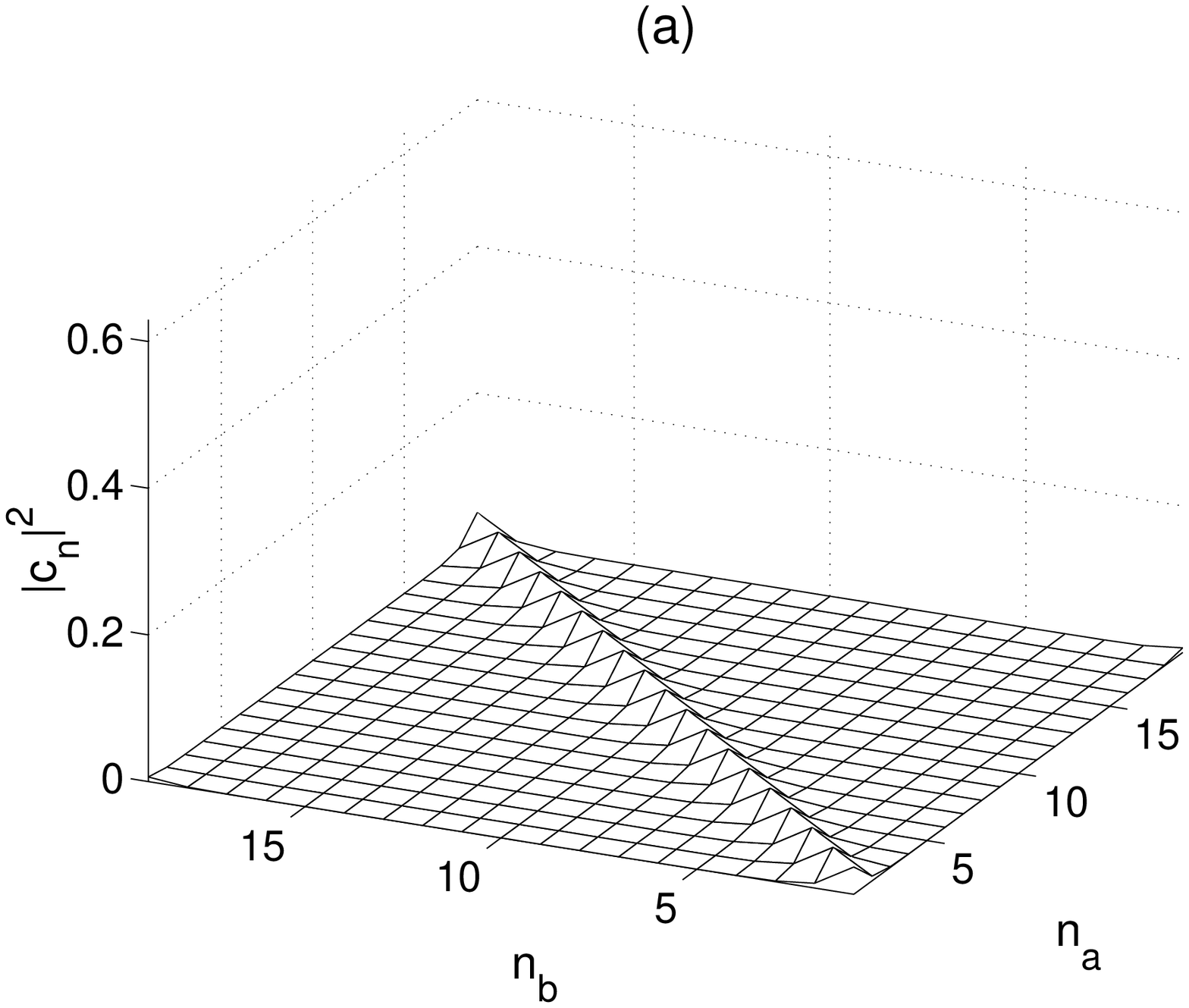}
    \includegraphics[scale=0.3]{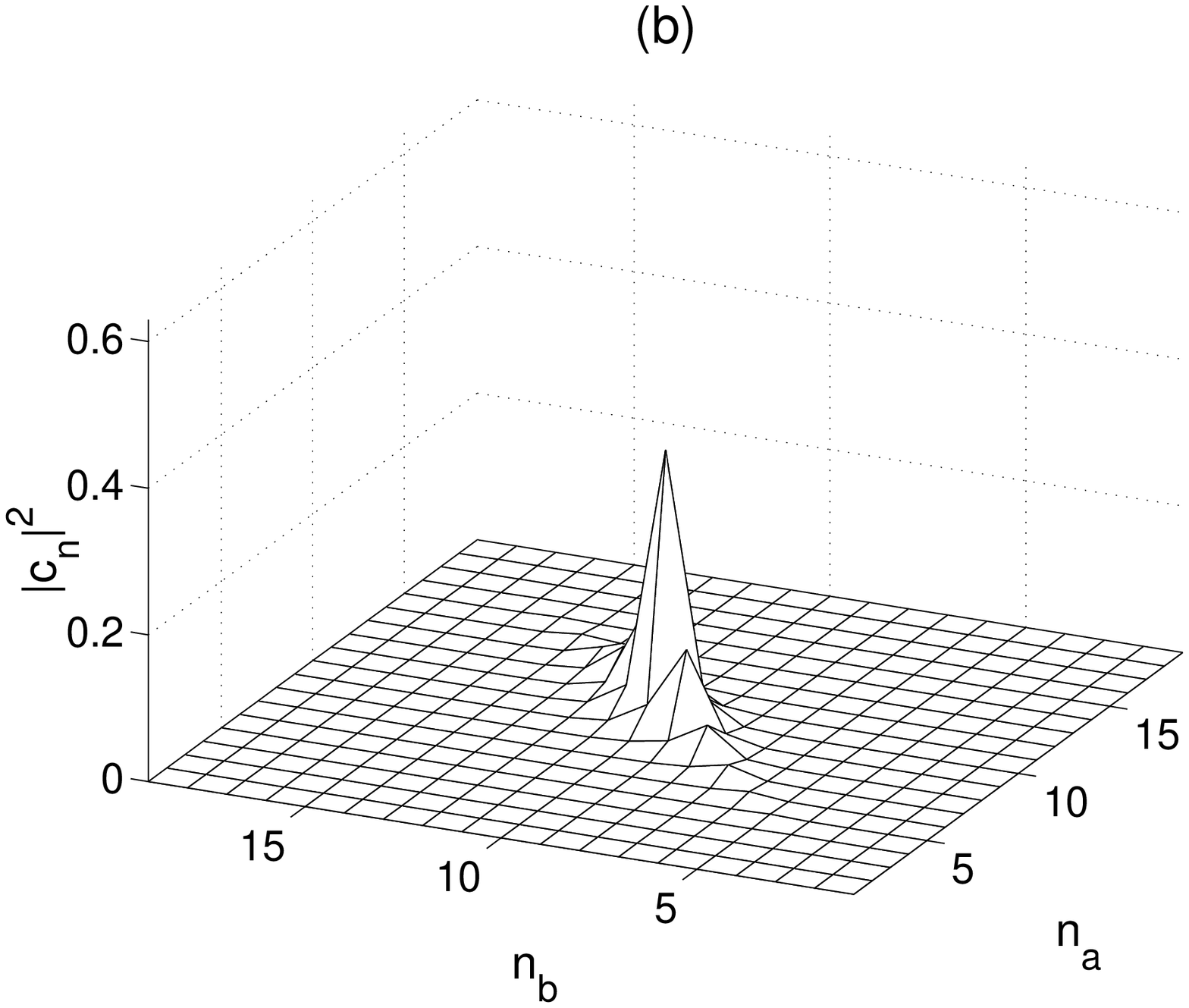}
    \includegraphics[scale=0.3]{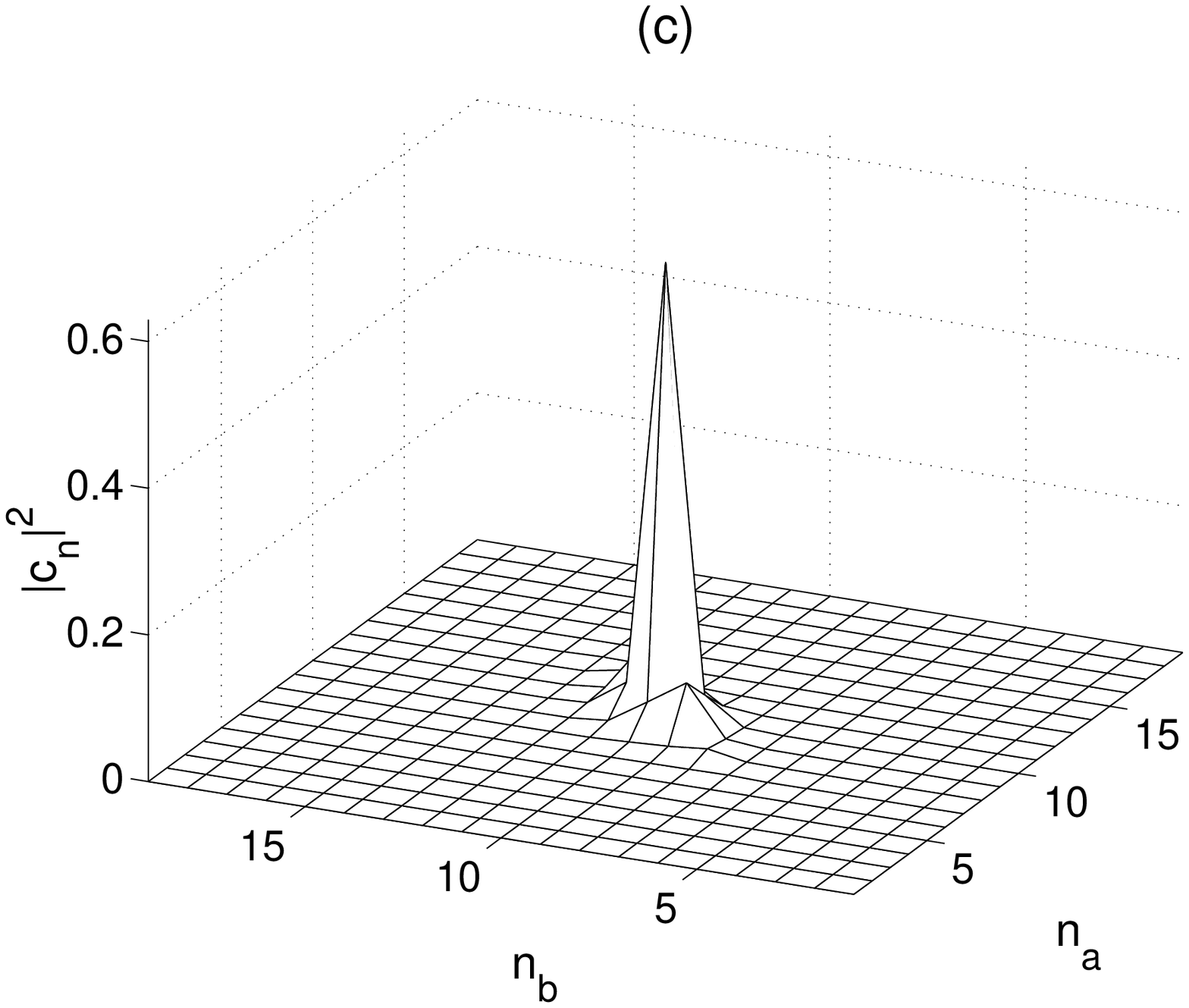}
    \includegraphics[scale=0.3]{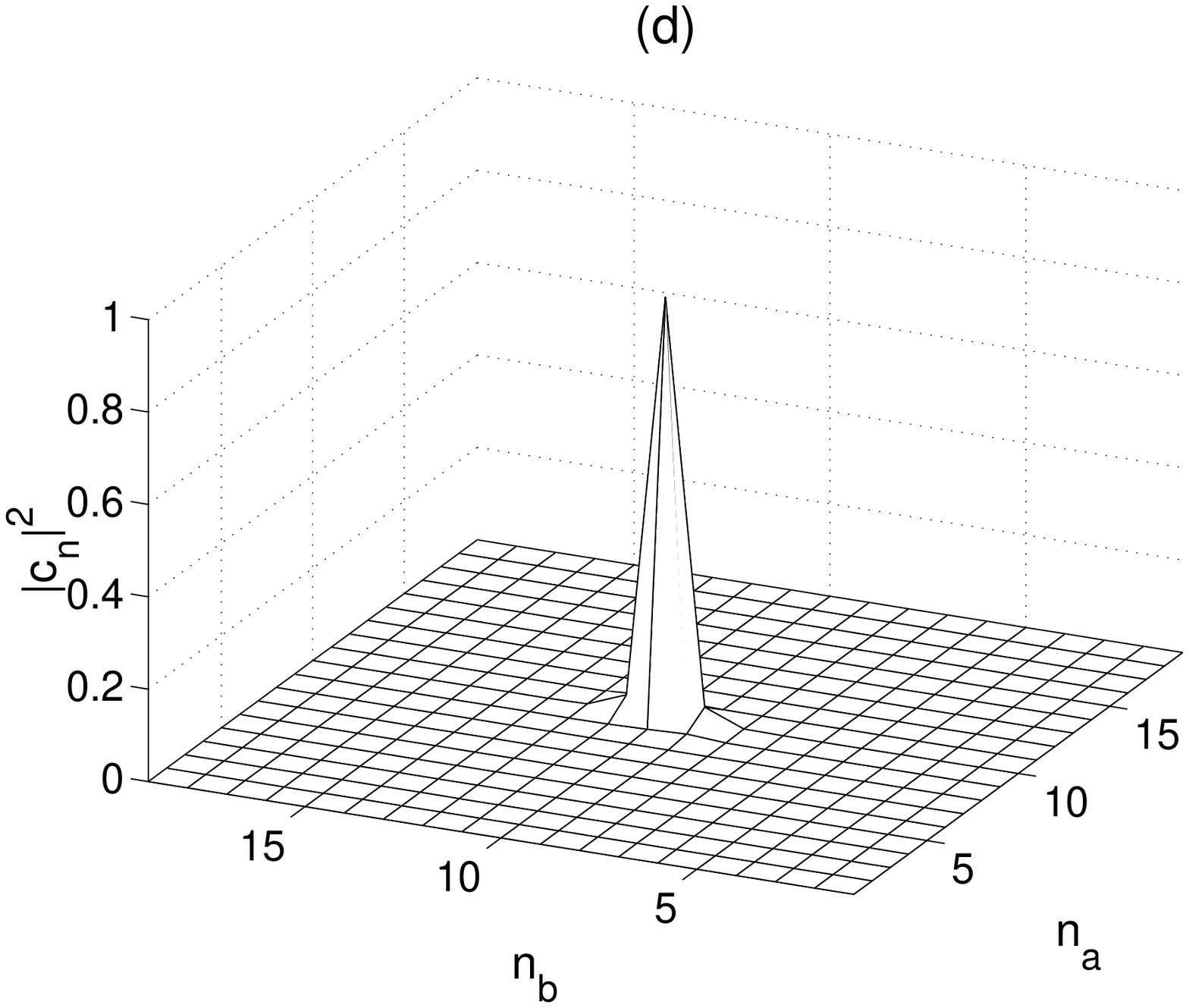}
  \end{center}
  \caption{Square wave function amplitudes $|c_n|^2$ corresponding to
    the ground state as a function of the positions of the two fermions
    along the chain $n_a$ and $n_b$. $f=19$ and $\gamma=4$.
    Point impurity at the site $\ell=10$. (a) Homogeneous chain, $m_e=1$. (b)
    $\gamma_{imp}=4.5$, $m_e=1$. (c) $\gamma_{imp}=5$, $m_e=1$. (d)
    $\gamma_{imp}=5$, $m_e=0.2$.}
  \label{fig4}
\end{figure}

If we analyze this ground state, we observe that as $\gamma_{imp}$
increases, the localization around the impurity increases too, as
shown in Fig \ref{fig4}. In particular, the main contribution to
the wave function corresponds to the bound states centered around
the impurity. There exists also a small contribution that
corresponds to states with particles in adjacent sites around
this local inhomogeneity.

\begin{figure}[h]
  \begin{center}
    \includegraphics[scale=0.4]{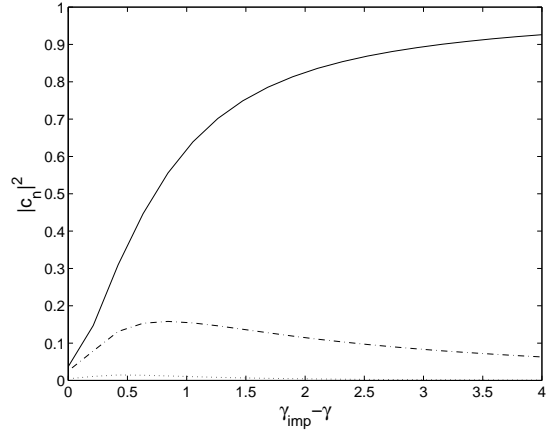}
  \end{center}
  \caption{Some components of the wave function corresponding to the
    ground state. $N=2$, $f=19$, $\gamma=4$ and $m_e=1$. Two particles centered on
    the impurity (continuous line).  Two particles in adjacent sites
    with one of them centered on the impurity (dashed--dotted line).  Two
    particles separated by one site and one of them on the impurity
    (dotted line).}
  \label{fig5}
\end{figure}

If we analyze the contribution  of the components of the wave
function of the ground state corresponding to the two particles
centered around the local inhomogeneity in the same site, in
adjacent sites, and separated by one site, as shown in Fig
\ref{fig5},  we observe that the localization increases very
rapidly with the magnitude of the impurity. Varying the value of $m_e$
from unity amplifies this effect even further.  We note that, in this
case, as harmonic terms are homogeneous (null), there exists no
Anderson--like localization.

\subsection{Localization in a twisted chain}
\begin{figure}[h]
  \begin{center}
    \includegraphics[scale=0.3]{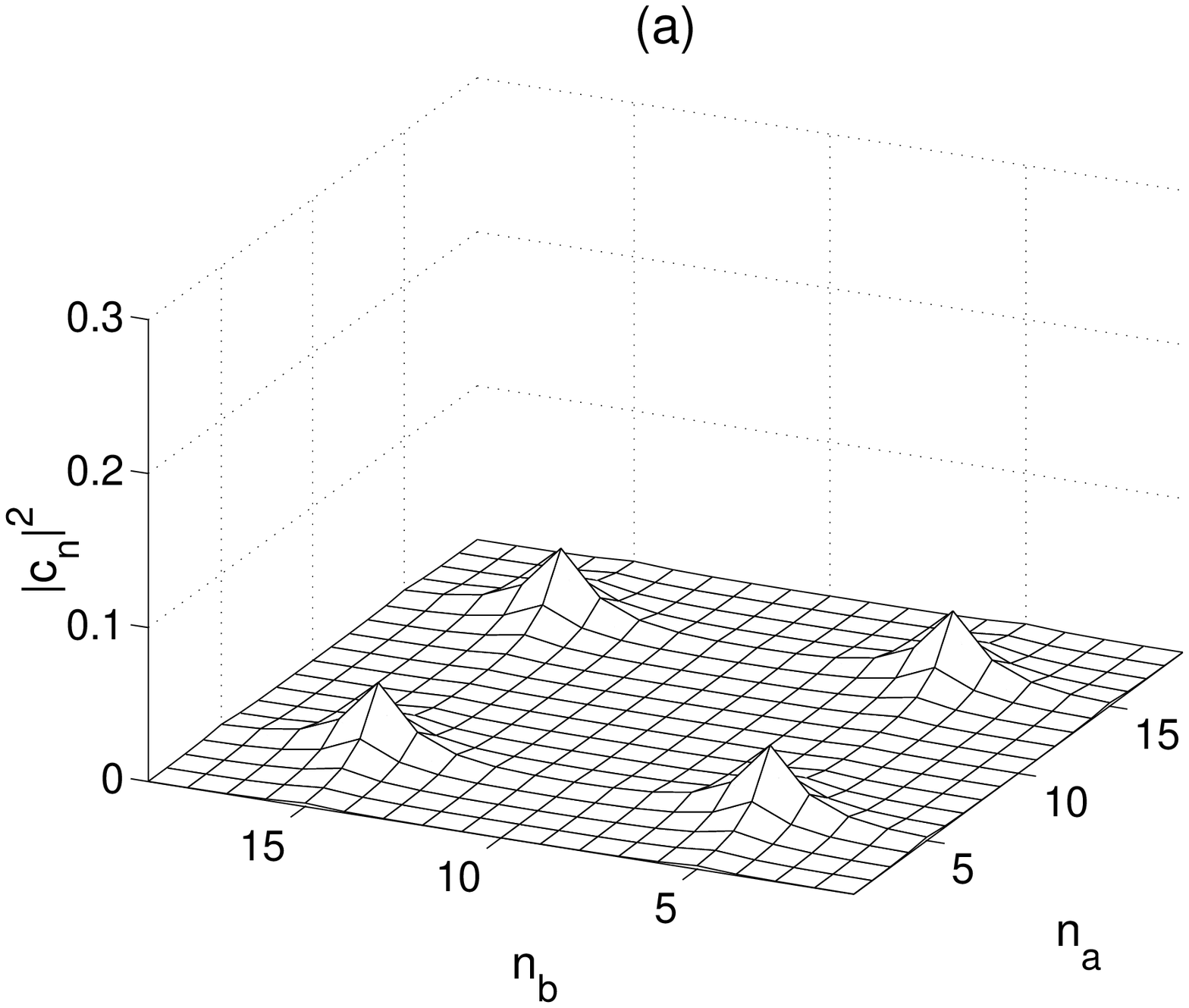}
    \includegraphics[scale=0.3]{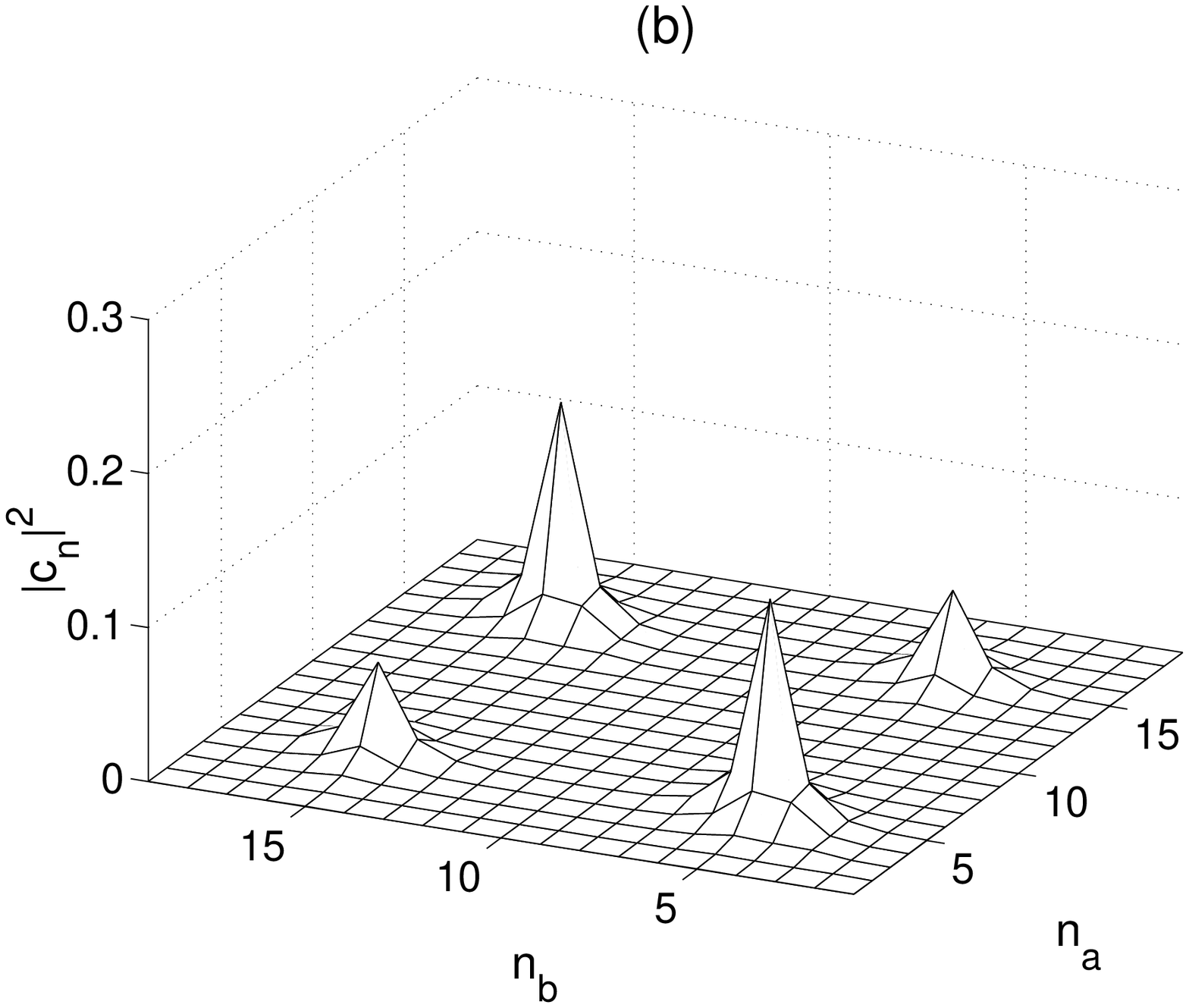}
    \includegraphics[scale=0.3]{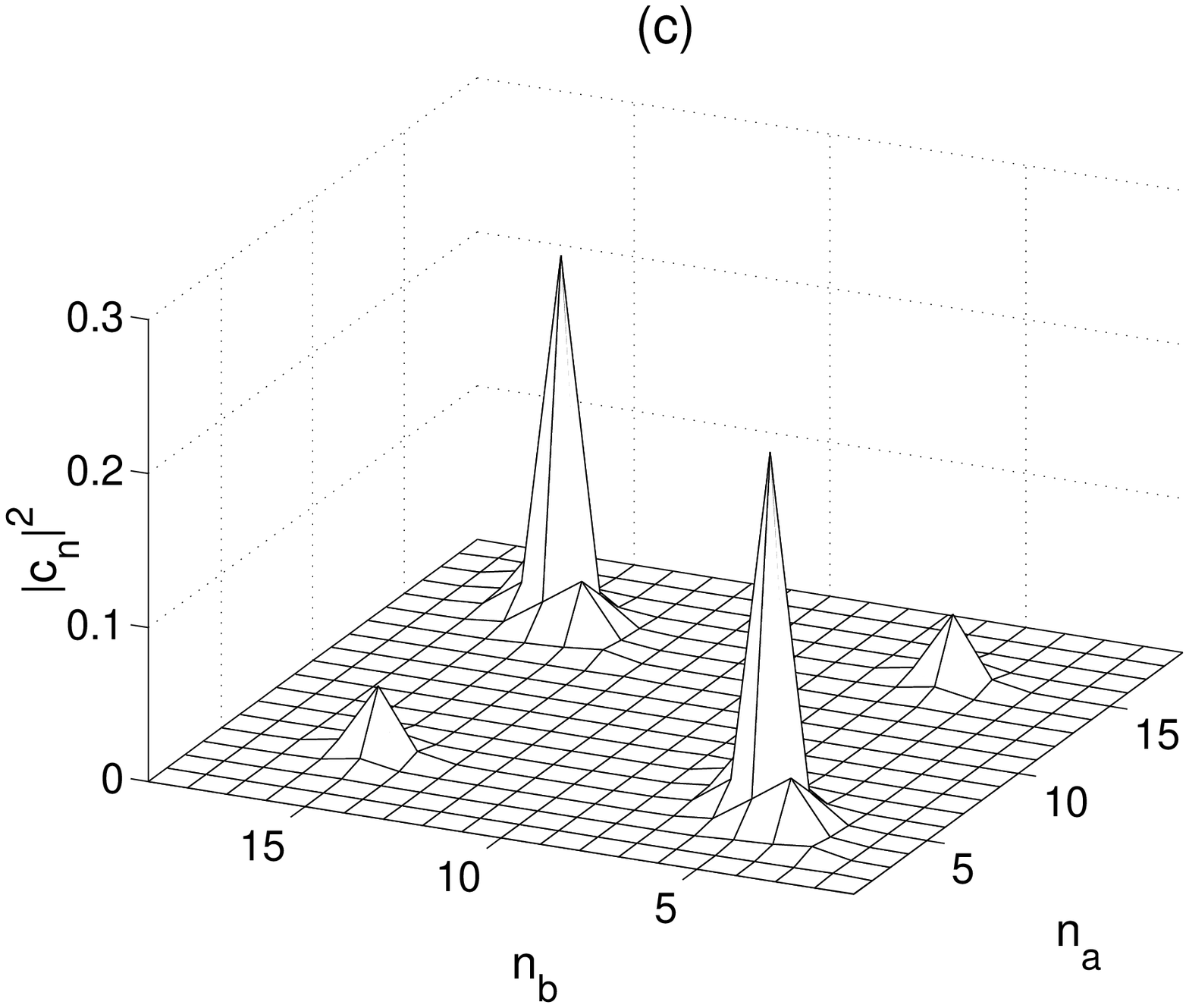}
    \includegraphics[scale=0.3]{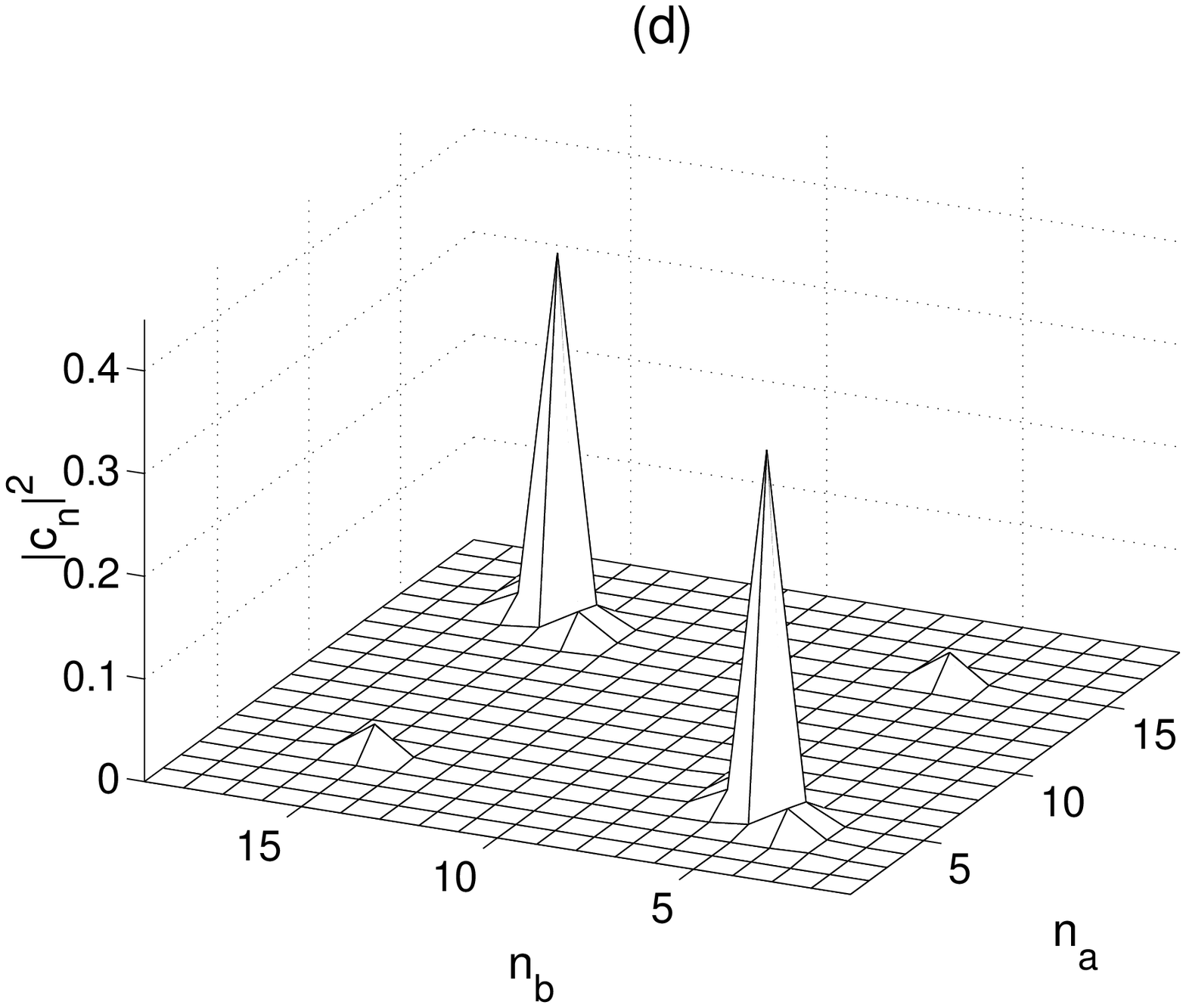}
  \end{center}
  \caption{Square wave function amplitudes $|c_n|^2$ corresponding
    to the ground state as function of the positions of the two
    fermions $n_a$ and $n_b$ along the twisted chain. Long range
    interaction between sites $m=5$ and $\ell=15$ with
    $\alpha_{m\ell}=1$ and $f=19$. (a) $\gamma=0$ (harmonic case),
    $m_e=1$. (b) $\gamma=2$, $m_e=1$. (c) $\gamma=4$, $m_e=1$. (d)
    $\gamma=4$, $m_e=0.2$.}
  \label{fig6}
\end{figure}

In order to simulate the twisted chain shown in Fig.\ \ref{fig2}a, we
consider a long--range hopping term between sites $m$ and $\ell$ given
by parameter $\alpha_{m,\ell}$. As Fig.\ \ref{fig6} shows, this
coupling generates a localized bound state around the sites $m$ and
$\ell$ that is a ground state of the system, a phenomenon similar to
that shown in Fig \ref{fig3}.
Although there exist some degree of localization in the harmonic
case ($\gamma=0$) due to an Anderson--like effect, the existence
of bound states due to the anharmonicity parameter $\gamma$
strongly increases the localization. Similar results have been
obtained with different values of the parameter $\alpha_{m\ell}$.

\subsection{Localization in a bent chain}
To simulate the bend shown in Fig. 2b, we introduce an additional term
that takes into account the interaction between the two neighbouring
sites of the vertex. In this case, if we suppose that the hopping term
varies as the inverse of the square of the distance between sites, the
parameter $\alpha$ can be related to the wedge angle $\theta$ through
$\alpha=\frac{1}{2}/(1-\cos\theta)^{-1}$.

\begin{figure}[h]
  \begin{center}
    \includegraphics[scale=0.3]{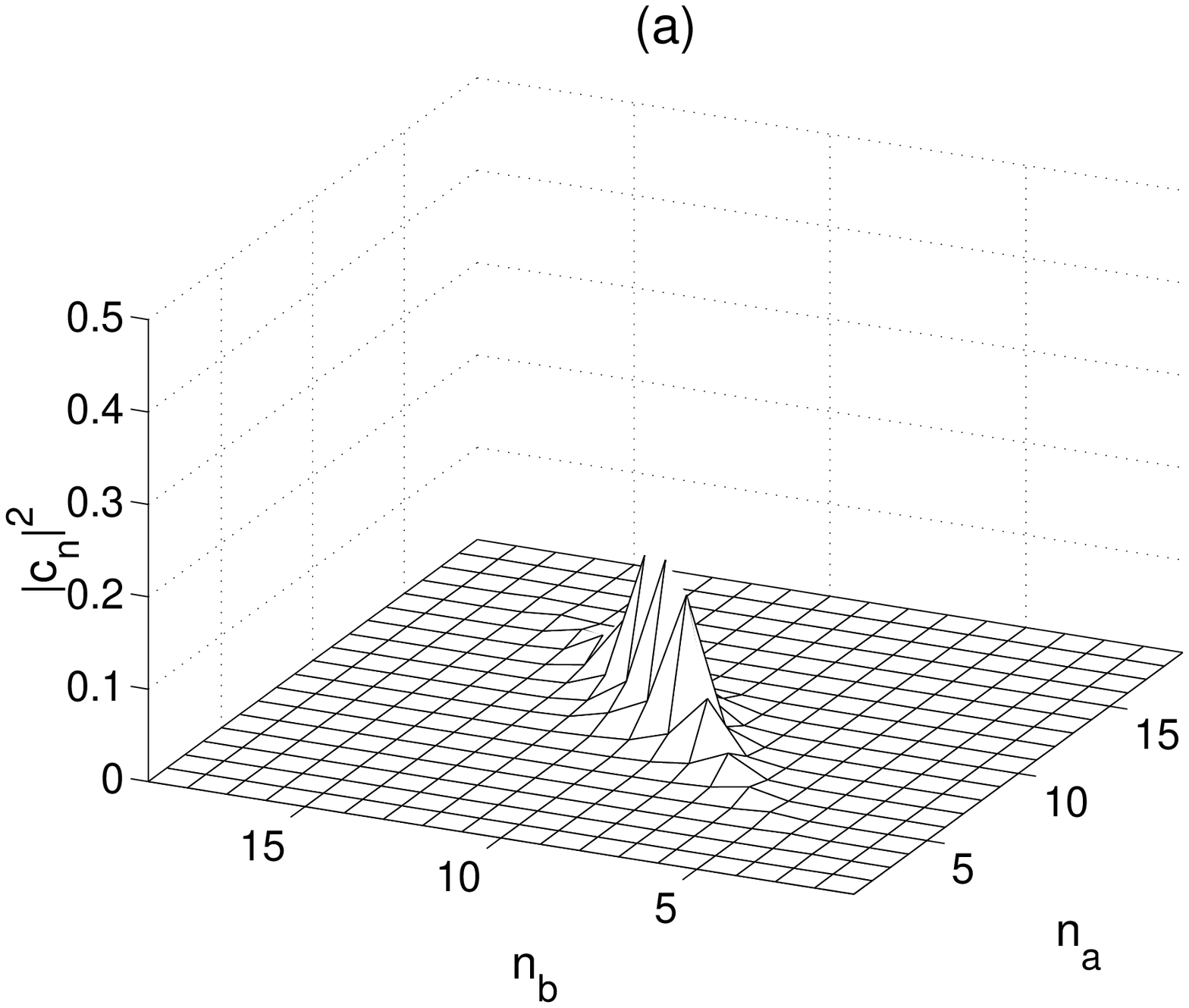}
    \includegraphics[scale=0.3]{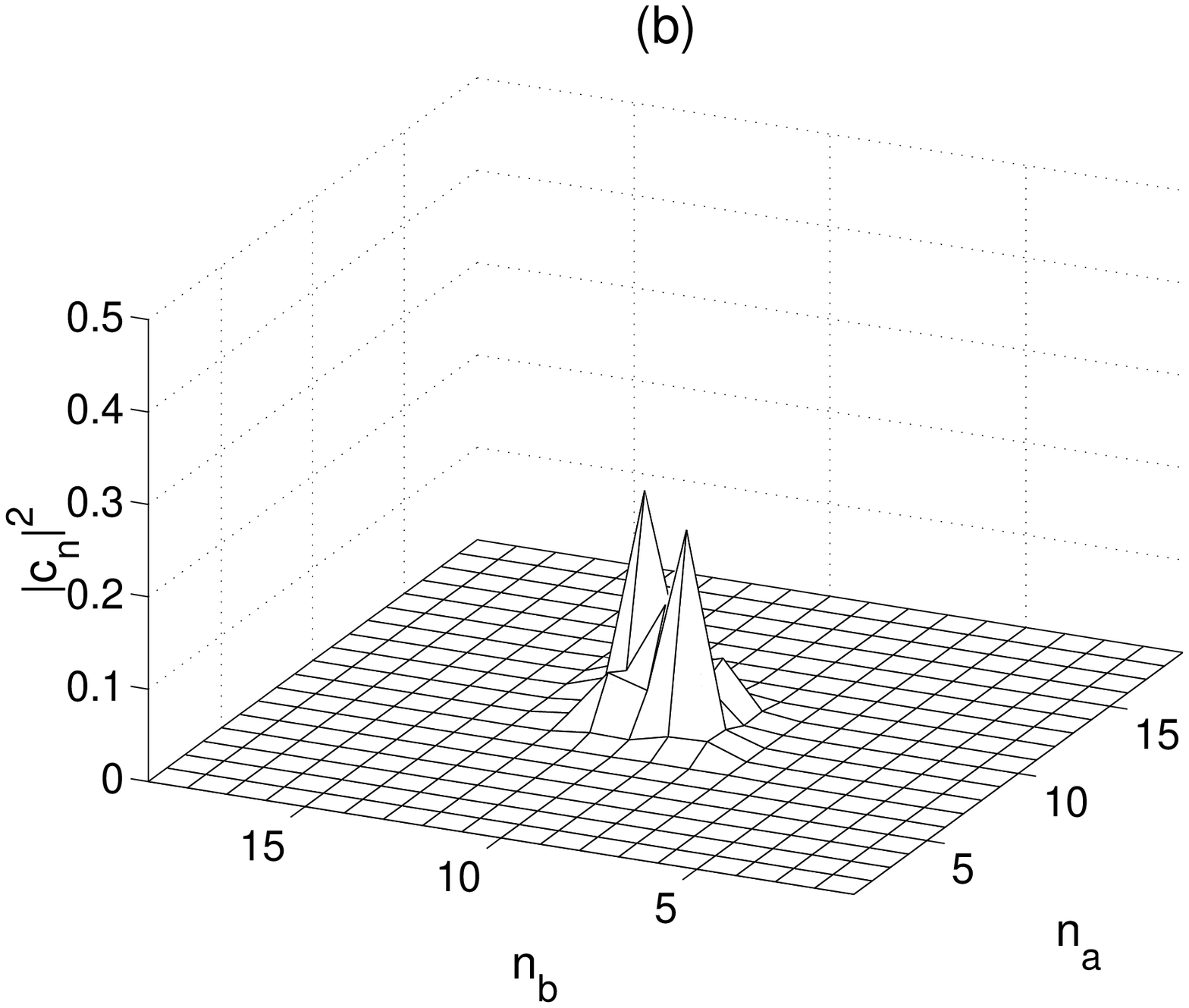}
    \includegraphics[scale=0.3]{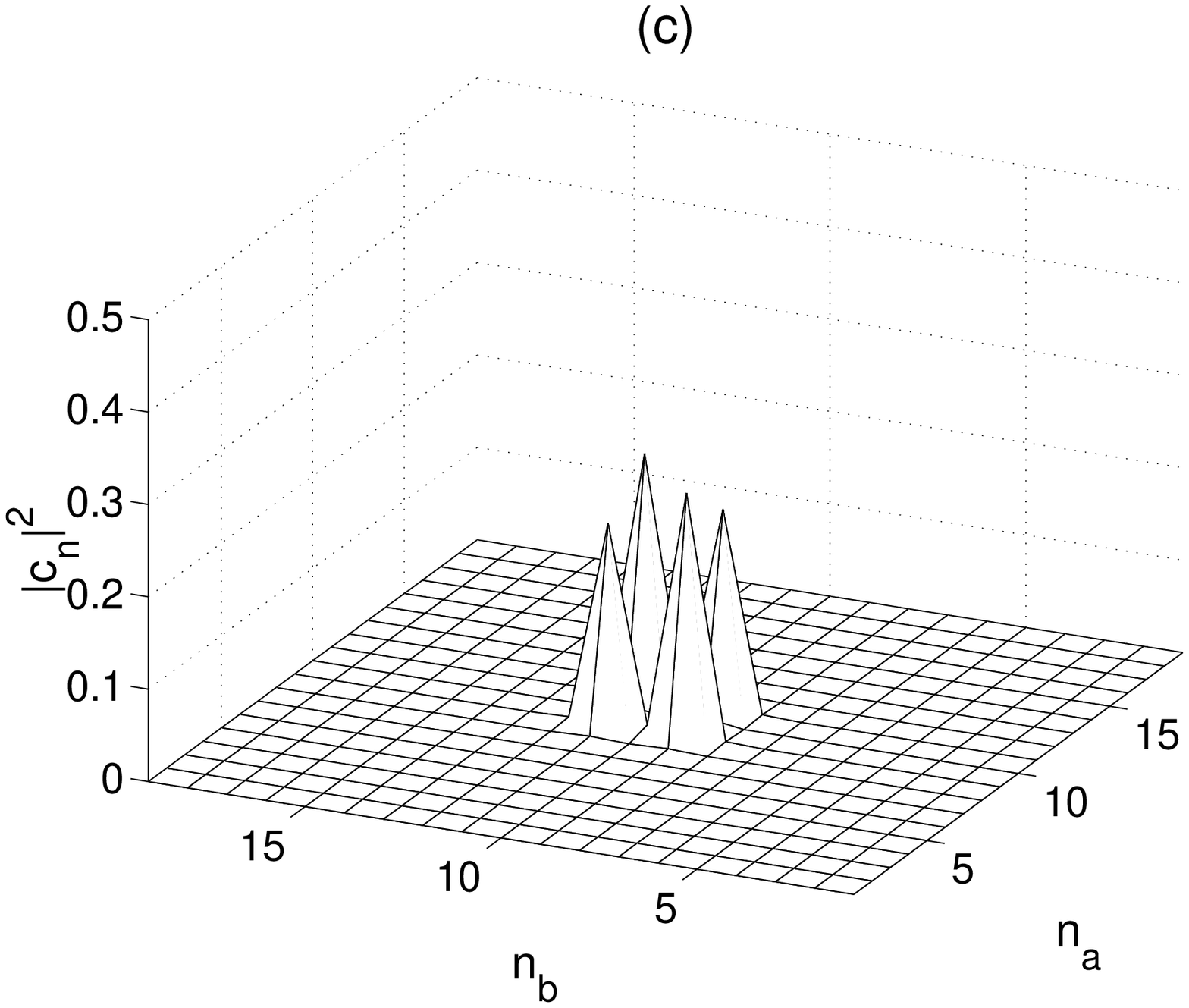}
    \includegraphics[scale=0.3]{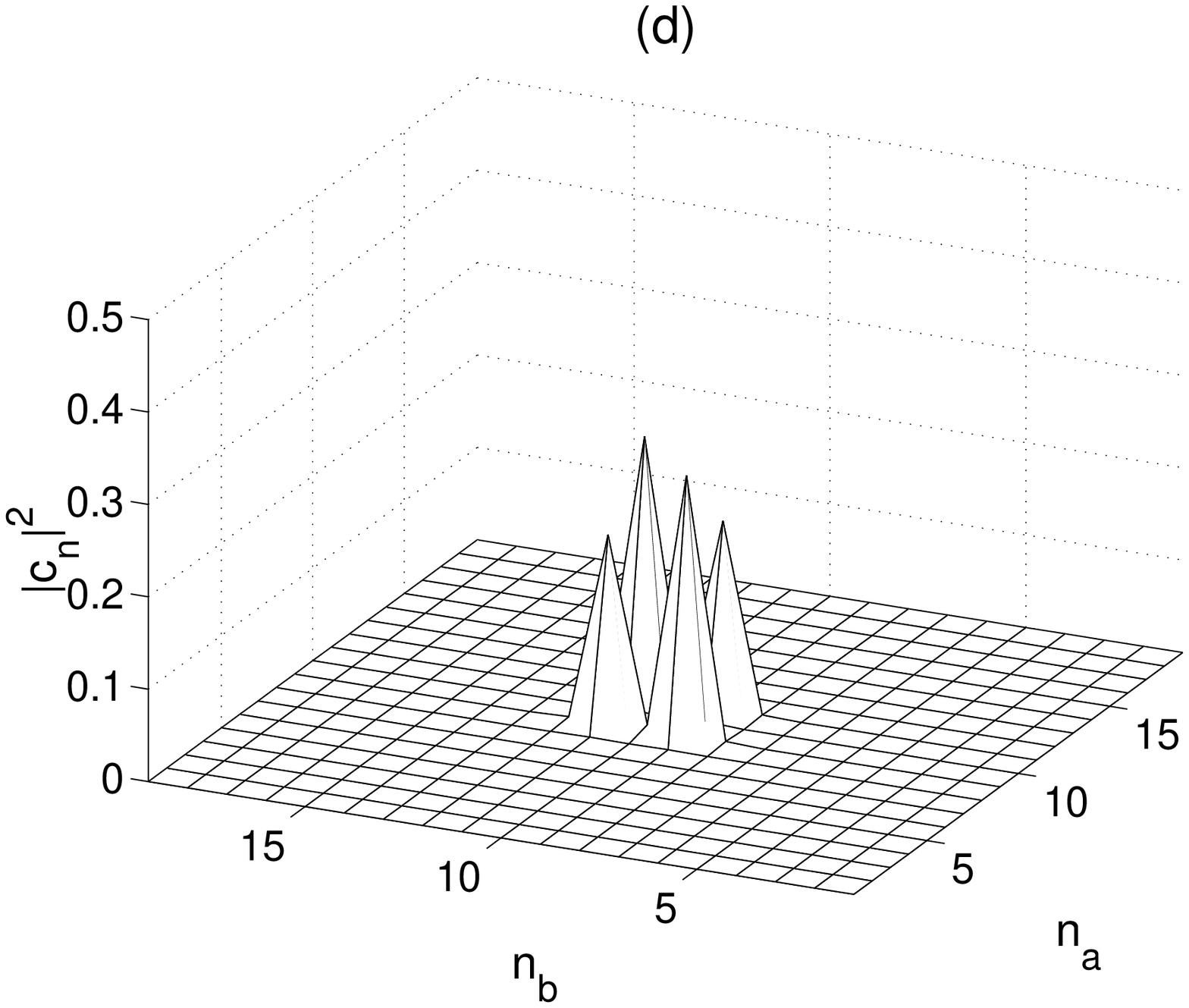}
  \end{center}
  \caption{Square wave function amplitudes $|c_n|^2$ corresponding
    to the ground state as a function of the positions of the two
    fermions $n_a$ and $n_b$ along the bent chain. $f=19$ and
    $\gamma=4$.  (a) $\theta=\pi$, $m_e=1$. (b) $\theta=\pi/3$,
    $m_e=1$. (c) $\theta=\pi/10$, $m_e=1$. (d) $\theta=\pi/10$,
    $m_e=0.2$.}
  \label{fig7}
\end{figure}

As shown in Fig.\ \ref{fig7}, due to the existence of this
long--range interaction, there exist a localization phenomenon
around the vertex of the chain. If the wedge angle is small
enough, the ground state is mainly a bound state with the two
particles localized in the neighbouring site of the vertex, but when
this angle decreases, the contribution of the components
corresponding to non--localized states with particles around the
vertex becomes significant. In the limit $\theta \rightarrow 0$,
the lattice becomes a T-junction. We have found that in this
system, the ground state is mainly localized around the junction.

\begin{figure}[h]
  \begin{center}
    \includegraphics[scale=0.4]{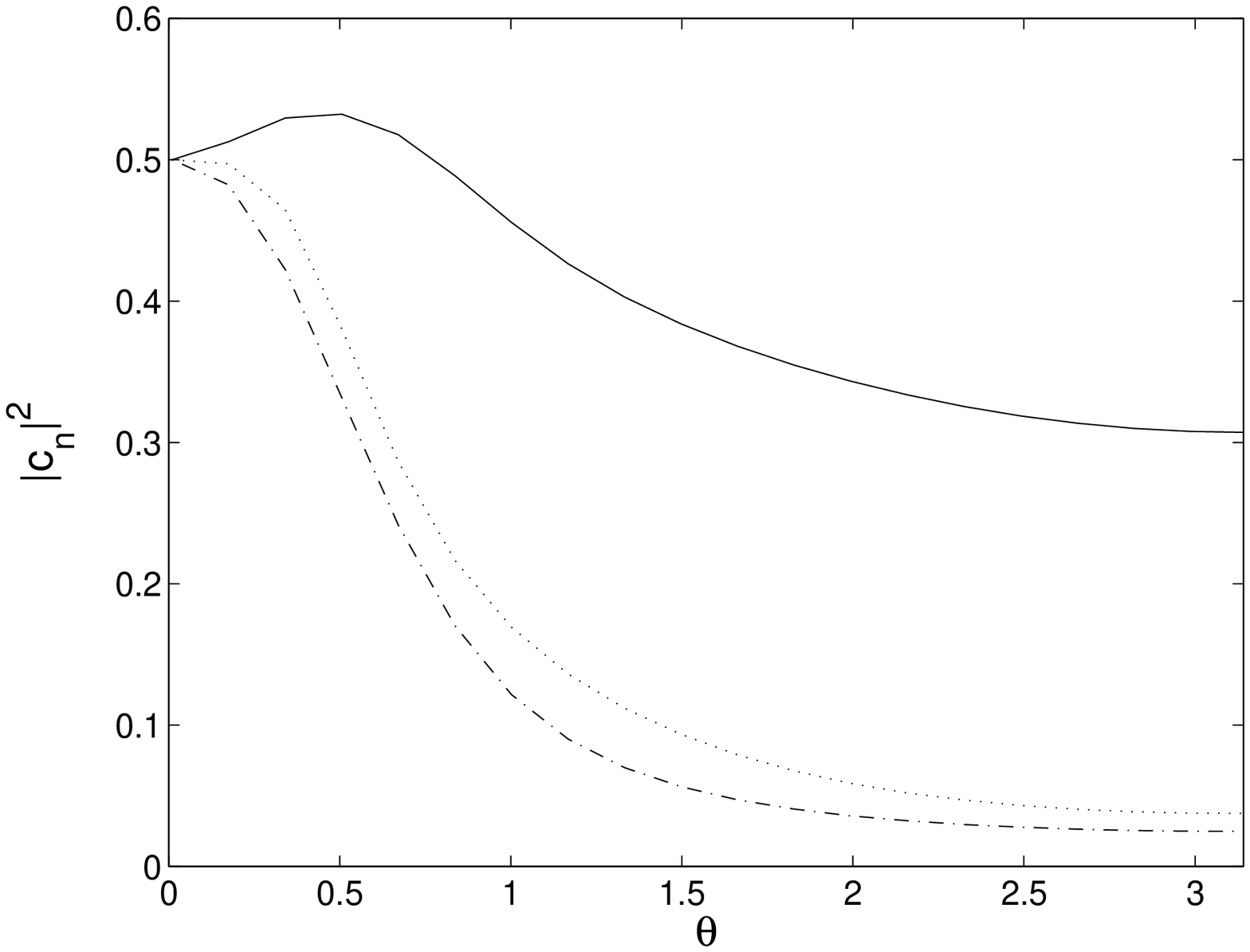}
  \end{center}
  \caption{Some components of the wave function corresponding to
    the ground state. $N=2$, $f=19$ and $m_e=1$. Localized state
    corresponding to the two particles at the neighbor site of the
    vertex and $\gamma=4$ (continuous line).  Two particles in a
    non--localized state at neighbouring sites of the vertex and
    $\gamma=4$ (dashed-dotted line).  Two particles in a non--localized
    state at neighbouring sites of the vertex and $\gamma=0$ (dotted
    line).}
  \label{fig8}
\end{figure}

We have compared this localization effect with the Anderson-like
localization in the harmonic system ($\gamma=0$). As shown in Fig.
\ref{fig8}, the existence of bound states in the anharmonic case
implies that the localization effect due to the curvature of the
system increases. This enhancement decreases when $\theta$
decreases, although there exists a maximum around $\theta\approx0.5$.

We note that this model, to give a more realistic approximation of a
bent chain, must be improved to take into account the long--range
interaction between all sites that becomes significant when the
angle $\theta$ is small enough. We have considered the model given
by the Hamiltonian
\begin{eqnarray}
\hat H & = & -\gamma \sum_{j=1}^{f}  a_j^\dag a_j b_j^\dag b_j
 -\sum_{j=1}^{f} \sum_{i>j} \frac{1}{d_{ij}^2}
( a_{i}^\dag a_{j}+a_{j}^\dag a_{i})
  - \nonumber \\ & & \sum_{j=1}^{f} \sum_{i>j}
  \frac{m_e}{d_{ij}^2}( b_{i}^\dag b_{j}+b_{j}^\dag b_{i}),
\label{Ham_fer_lr_t}
\end{eqnarray}
where $d_{ij}$ represents the distance between sites $i$ and $j$.
We have found the same qualitative behavior.

\section{Higher number of quanta}

In previous sections, we have restricted our studies to the case
$N_a=N_b=1$. Proceeding as the same way, it is possible --in
principle-- to construct the Hamiltonian matrix for any value of
the quantum numbers $N_a$ and $N_b$ and to calculate the spectrum.
However, the computational effort increases rapidly and can go
beyond the limits of computational convenience. Nevertheless, we
have studied some cases involving a higher number of fermions. In
particular, we have considered the case $N_a=2$ and $N_b=1$ and
the case $N_a=N_b=2$.

In general, we have found the same qualitative behavior than in
the previous case. In the homogeneous system, if the anharmonic
parameter is high enough, the ground state is mainly a localized
state, in the sense that there exists a high probability to find
two different fermions at the same point of the lattice, but due
to the translational invariance of the system, with equal
probability of finding these two particles at any site of the
system. However, we observe that the main components of the ground
state correspond to states where fermions of the same type are as
far apart as possible from each other. This is a similar
effect as due to the finite--size of the chain where the ground
state is weakly localized around the center of the chain. When a
fermion is close to other of the same type, the hopping in that
direction is limited, as in the case of a finite--size chain.

\begin{figure}[h]
  \begin{center}
    \includegraphics[scale=0.3]{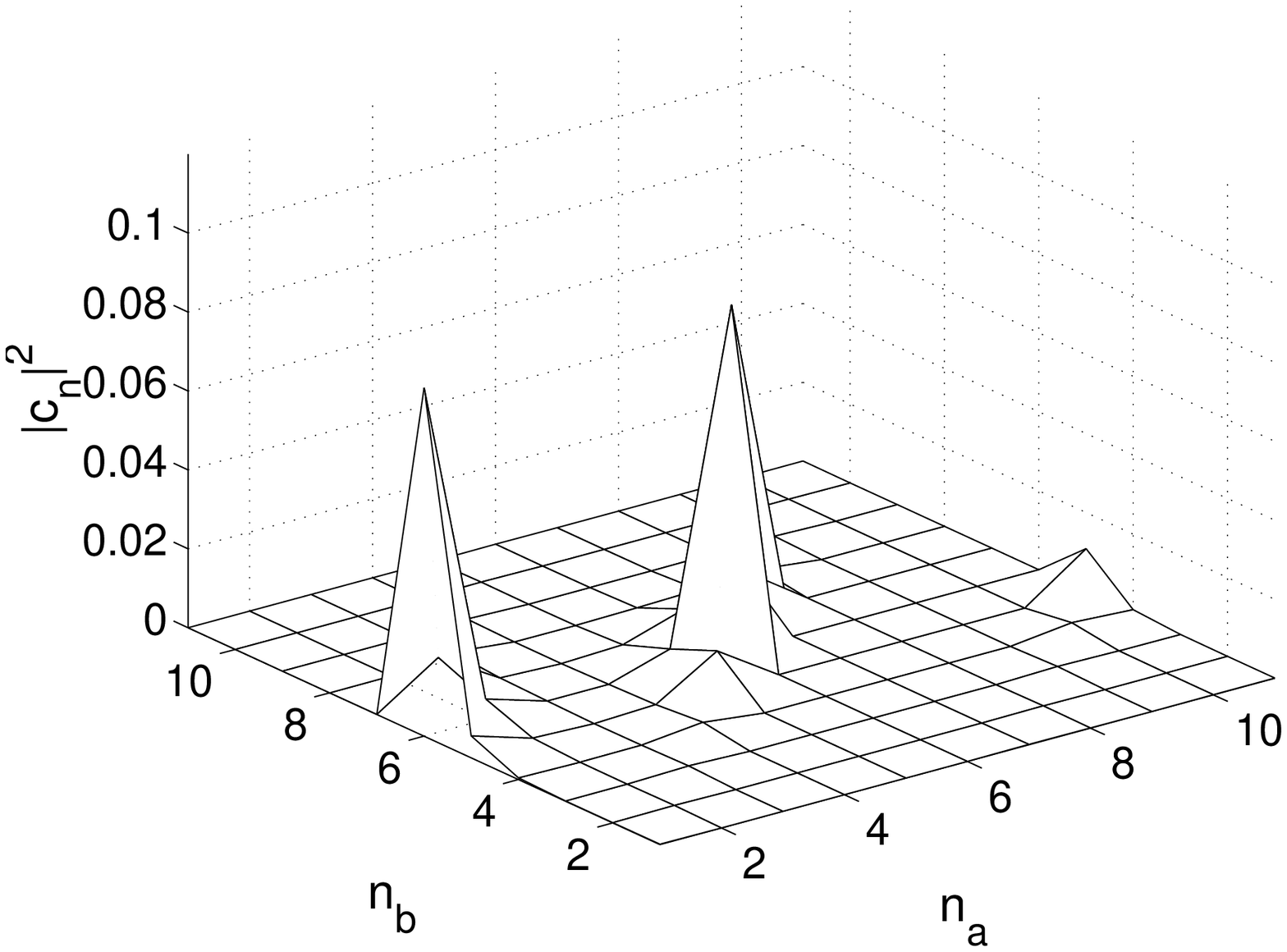}
    \includegraphics[scale=0.3]{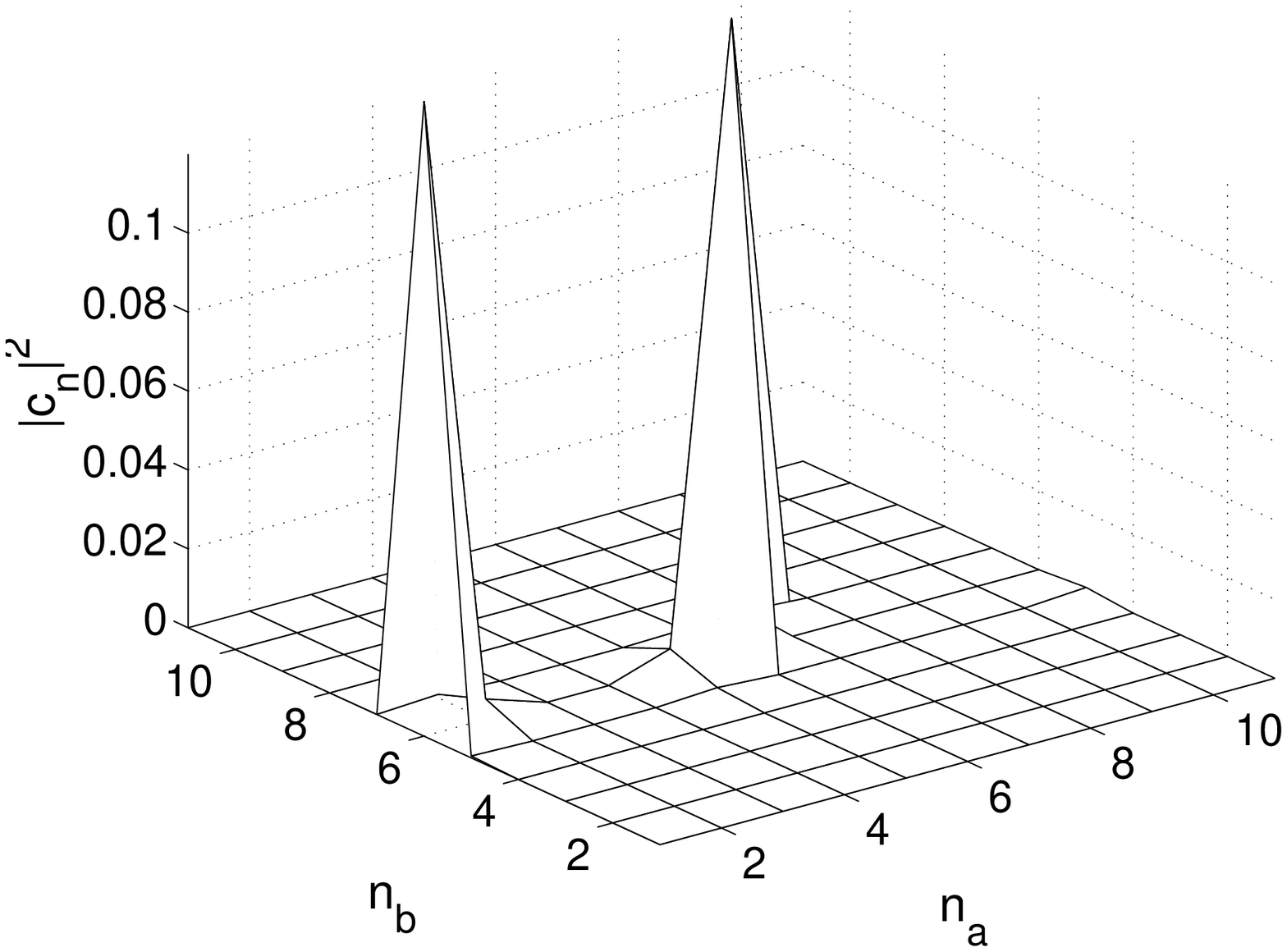}
 \end{center}
  \caption{Some components of the wave function of the ground
    state corresponding to the case $f=11$, $N_a=2$, $N_b=1$,
    $\gamma=4$, point impurity at the site $\ell=6$ and $\gamma_{imp}=5$.
    $n_a$ represents the position of one of the fermions of type (a)
    (the other fermion is located at $n_a+5$) and $n_b$ the position
    of the fermion of type (b).  $m_e=1$ (left), $m_e=0.2$ (right).}
  \label{fig9}
\end{figure}

When we introduce some local inhomogeneities in the system, we
have observed similar localization phenomena as noted above.
In Fig \ref{fig9}  we show the case $N_a=2$ and $N_b=1$ with a
point impurity at the anharmonic parameter. We observe that the
ground state is mainly a bound state. The two different fermions
are mainly in a localized state centered at the impurity with the
other fermion in the other extremum of the chain. We note that
there exists a significant contribution of other components
corresponding to  localized states of the two different fermions
in the impurity and the other one in different sites of the chain,
this contribution being more significant when it corresponds to
states where the two fermions are separated by a large number of
sites. This system, in the context of excitons in ring geometries,
is usually called ortho--trion, and can be viewed as an exciton
plus an additional electron smeared over the ring \cite{Ro01}.

In the other cases, when a local inhomogeneity is introduced by
means of a  long--range interaction term, or  $N_a=N_b=2$ is
considered, the behavior is similar. The ground state corresponds
to a localized state centered at the local inhomogeneities where
different fermions are together and fermions of the same type are
located as further apart as possible one from the other.

\section{Conclusions}

In this work we have shown some results related with the existence
and properties of quantum breathers in a fermionic Hubbard model
with two kinds of particles of opposite spins. We have studied the
existence of localized states due to the nonlinearity and to the
influence of local inhomogeneities in these localized states. In
particular, we have found that these local inhomogeneities, due to
the geometrical factor and to a long--range interaction or an
impurity in the anharmonicity parameter, break the translational
invariance of the system and localize the ground state around a
particular site of the chain. We expect that these results are
rather general, and could be extended to a great variety of
systems.

\section*{Acknowledgments}
The authors are grateful for partial support under the LOCNET EU
network HPRN-CT-1999-00163. F. Palmero thanks Heriot-Watt
University for hospitality, and the Secretar\'{\i}a de Estado de
Educaci\'on y Universidades (Spain) for financial support.

\end{document}